# Status of the Computing for Research in Africa

## Ghita Rahal

CC-IN2P3, CNRS, France

May 20st, 2022

## Introduction

Research in any Science needs nowadays strong computing services to extract results and make discoveries.
What we define as computing service might rank from the underlying structure, namely networks, computers, storage, to applications and software but, as well, new techniques such as Artificial Intelligence to extract the expected results.

In order to estimate the overall needs in this field, we have launched a survey including all the people that we could reach, participants in ASFAP[1] as well as the attendants to the 2nd African Conference of Fundamental and Applied Physics ACP2021 that was held in mars 2022. Results of the survey can be found here[2].
In this paper we summarize the answers that were provided to the different questions of the survey and extract some general observations. Possible guidelines and recommendations to improve the situation are drawn in the conclusion.

## 1. Panel distribution

175 people filled in the survey out of which 167 were African citizens. 26 countries of the African continent were represented.
82% of the African citizens are based in Africa, the rest is what is defined as the diaspora, i.e., people that are based in other continents.
For the Non-African citizens, the motivation for participation was mainly an already established collaboration with African colleagues or students.
Figure 1 highlights the job situation of the participants: 48% are students and 39.4% hold a position in academia, research, engineering. More than 88% work in Africa, whether it is in their own country or in another African country. 10.9% reside out of Africa.

---

[1] ASFAP is the African Strategy for Fundamental and Applied physics: https://africanphysicsstrategy.org
[2] The result of the survey as of June 2022 is here:
https://twiki.cern.ch/twiki/pub/AfricanStrategy/AfComputing4IR/ASFAP:_Comp4IR_Survey.pdf

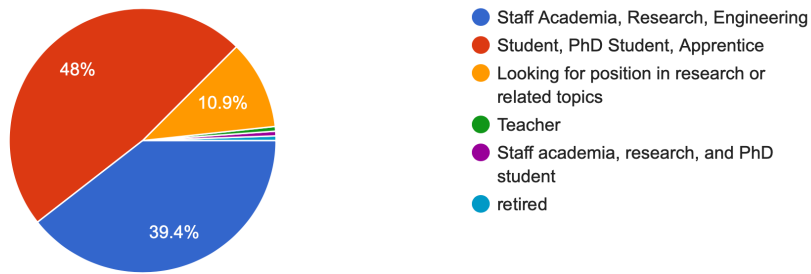

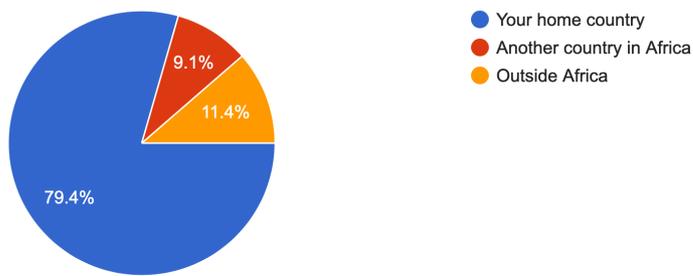

**Figure 1: shows the work situation of the participants (top) and the location of the activity (bottom)**

## 2. Field of Research

The field of research in which the participants are working is spread among many disciplines as shown in the figure below. The name of the fields is the name of the various working groups that are represented in the ASFAP. A large number of fields are concerned by the computing, fundamental physics such as Astrophysics and cosmology as well as physics domains such as the energy (biomass, fossil, nuclear and solar, etc.).

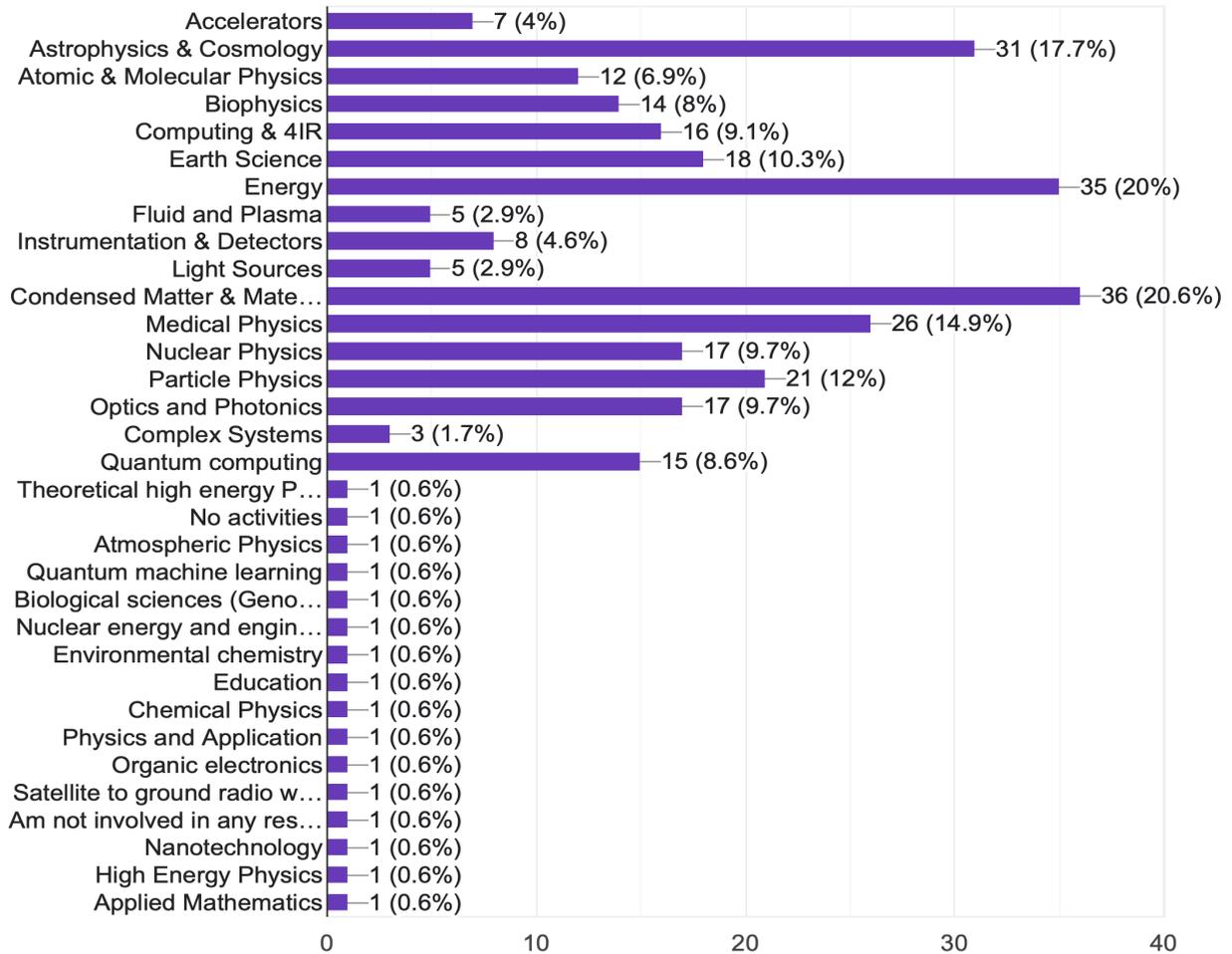

Figure 2: Field of Research covered by the participants

## 3. Properties of the exploited data sample

Depending on whether the scientist is part of a collaboration or working alone, the storage location and the volume of the data that he is analysing in order to extract results varies a lot. Figure 3 below shows the distribution and the magnitude of the sample.

- About 40% of the participants use data coming from an experiment or a collaboration they belong to, 15% use some open data from literature and about 35% use personal data. The numbers quoted here have been obtained summing up the numbers in the pie of fig 3 that fall in the same category.

- Considering the magnitude of the data samples, the majority of the samples are at the Gigabyte and Megabyte level. It is to be noted that the samples of Terabytes- and Petabytes-level are most of the time stored internationally.

- As to understand where the users are running their job to exploit the data, another graph not shown here shows that 57% of the scientists do it in their own laboratory or in their own country, while 20% do it internationally.

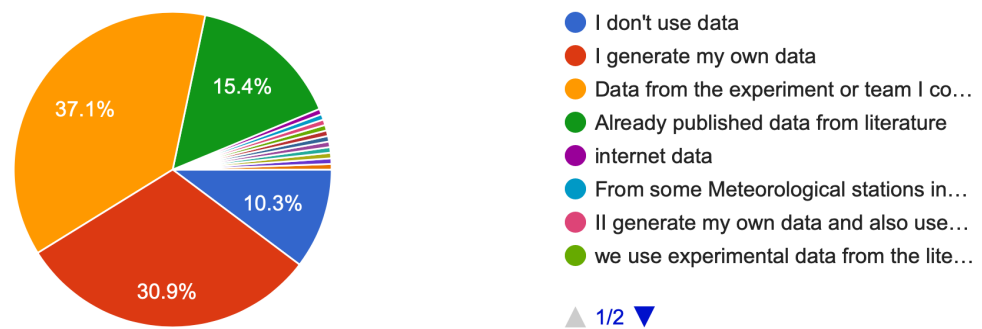

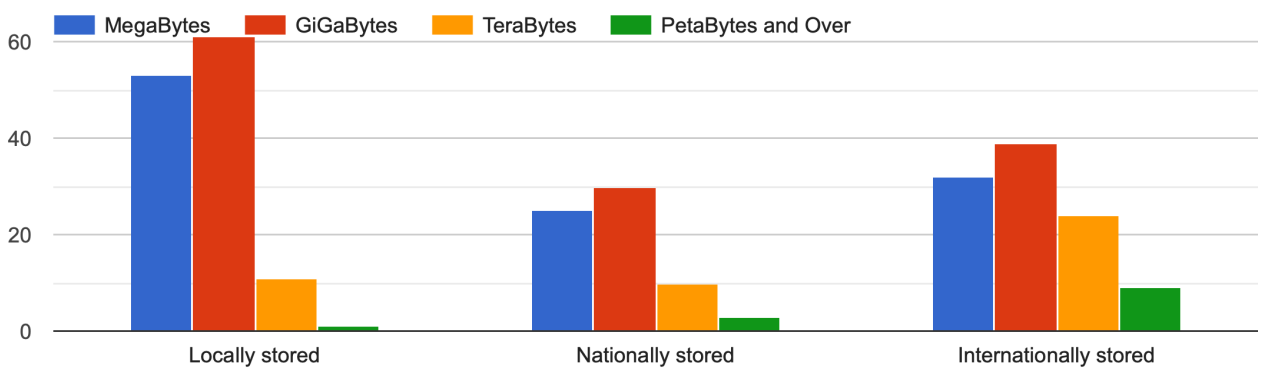

**Figure 3: upper graph shows the origin of the data that is exploited. The lower graph the magnitude of the sample as a function of the storage location**

## 4. Software and Tools to exploit data

Figure 4 below illustrate the type of software scientists are using:

- In the top graph of fig 4, we observe that 52.9% of the scientists use collaboration software and 48.4% use commercial software to exploit their data. Only 24.6% of them use exclusively their own software. It is important to note that 11.6% cannot use software because of lack of computing resources. Some comments collected below in this survey point out the fact that sometimes, the need is very modest (1 laptop for example) but even that is not available. Even if it has not been studied in this survey it might be worth investigating why even the modest requests are not satisfied.

- In the pie of the bottom graph of fig 4, usage of Artificial Intelligence (AI) or Deep Learning (DL) is represented. Already 20.9% of the scientists have introduced it in their toolbox. What is striking is that more than 73% of the scientists would like to use it but are prevented to do so either because they cannot find the information and/or training they need, or because they lack computing resources to execute it. This is important to keep in mind because of the growing importance of AI in many fields of sciences.

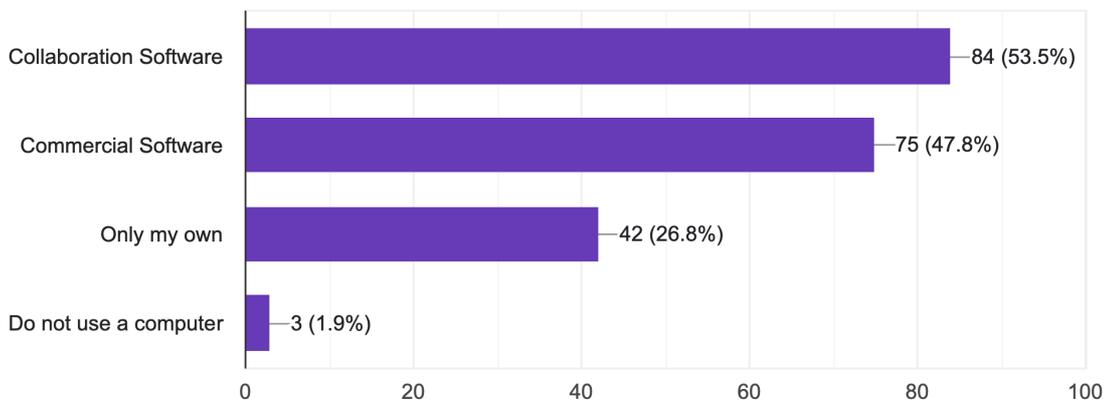

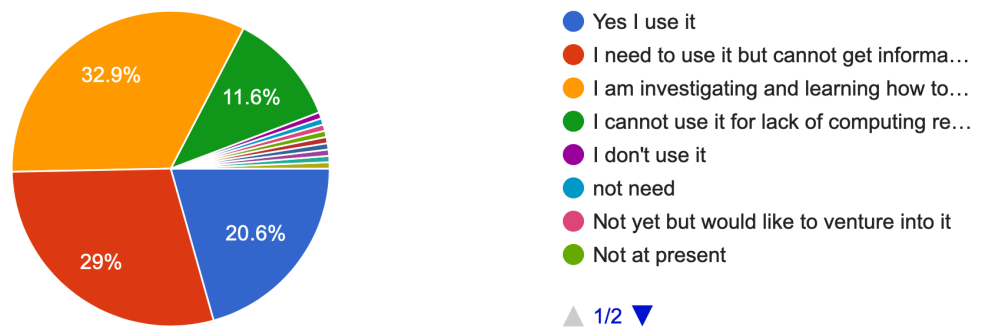

**Figure 4: Top is distribution of the type of software that is used by the researchers. Bottom shows the usage of Artificial Intelligence or Deep Learning by the scientists.**

## 5. Status of the infrastructure and tools

We have questioned the participants about the resources or knowledge that need to be provided in order to be able to use the magnifying effect of the computing to extract research results. They rated the different points in the figure 5:

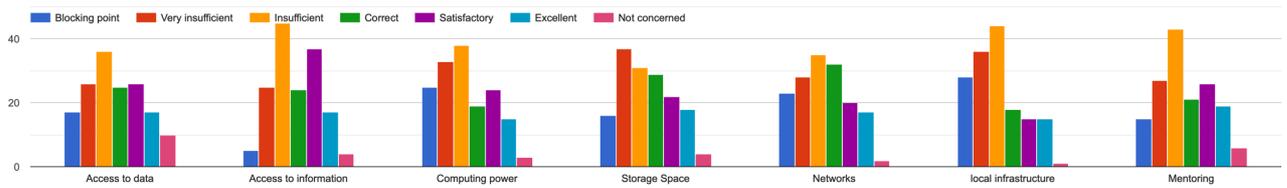

**Figure 5: Main blocking or satisfactory points to be able to exploit data:**

If we summarize the results they are as follows:

- **Access to data and access to information**: about half of the people experience blocking points leading to insufficient access to the data they need. This could be due to different reasons some of which could be equipment, networks or the fact that scientists feel isolated from the rest of their community.

- **Hardware resources, Computing power, Storage space and Networks**: 55 to 60% of the participant find it insufficient. When it comes specifically to local infrastructure, 66% of the people find it insufficient. The reason of the difference between the 2 percentage above is certainly that some people use international infrastructures that are more efficient: 20% of the people are based outside their home country and about 40% claim that they use resources abroad (see fig.1).

- **Mentoring:** training, guidance, lectures, etc..: 54% of the participants don't find it sufficient. This point will be detailed later.

## 6. Education and knowledge

Figure 6 below highlights one of the major problems of the education in sciences:

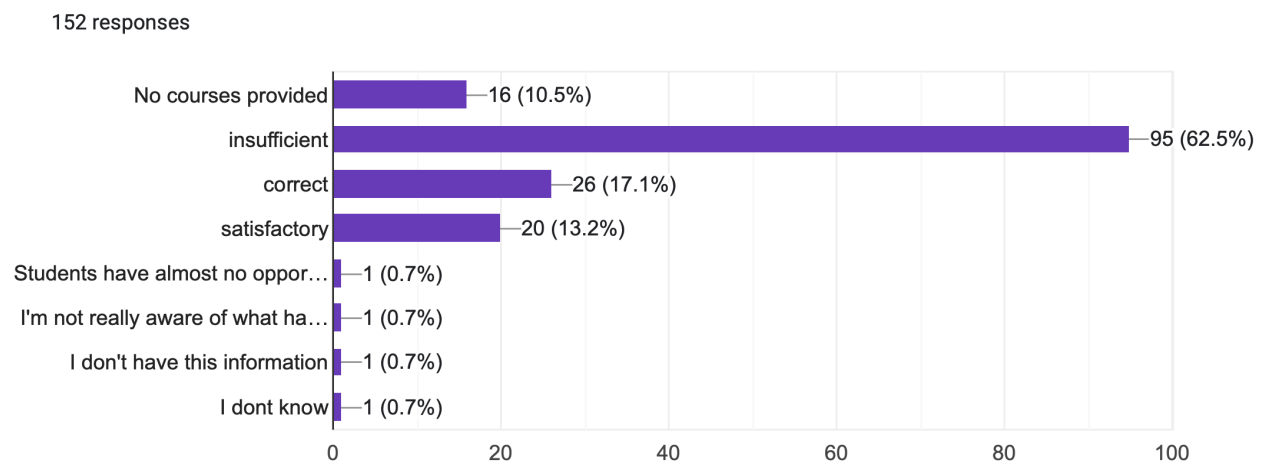

**Figure 6: Rating of teaching and training provided to students during their cursus**

**75%** of the participants claim they are provided no or insufficient level of courses or trainings during their cursus. This impacts not only the students themselves but it creates a generation of managers and scientists that are not made aware of the huge potential that computing can provide to their science field.

## 7. Bottlenecks

In Figure 7, we specifically question the users about the main bottlenecks there are facing when they want to use computing.

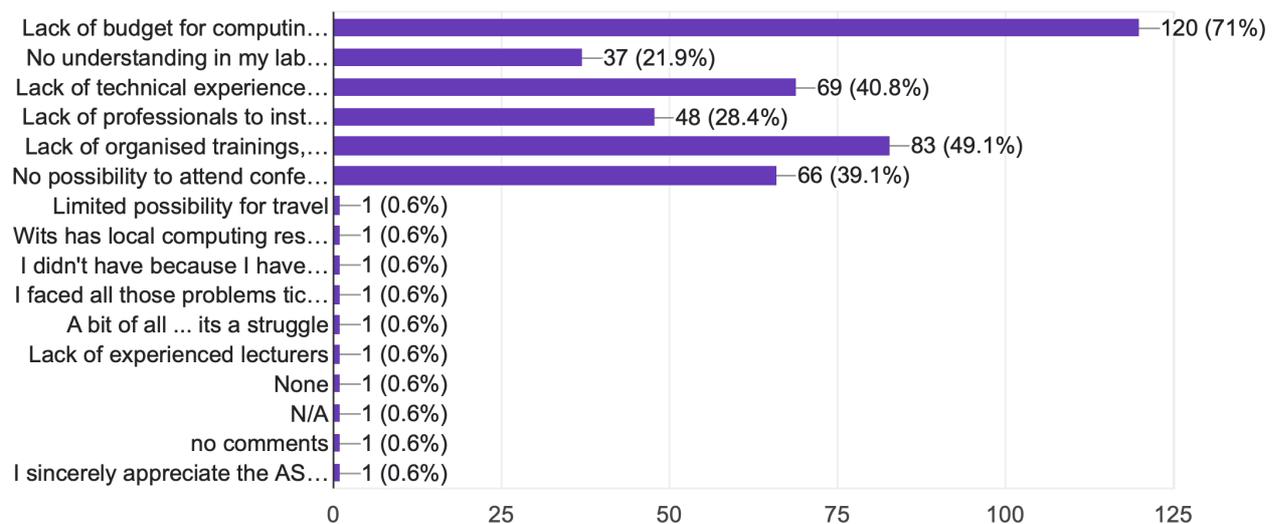

Figure 7: Problems faced by the scientists to perform their job

Two main group of answers:

- **Computing resources**: the largest number of responses stress the lack of budget for computing, the lack of technical support and the fact that the hierarchy and the stakeholders do not understand the need of computing for research: what would a scientist do with a computer when doing crystallography for example?

- **Education, knowledge**: the participants point out the lack of organised trainings and workshops and the difficulty to attend those organised abroad. More detailed information is found in the previous paragraph and fig 6 about the teaching and education in computing: 75% of the scientists are not provided courses and lectures, or at an insufficient level.

## 8. General comments from the survey

In the free comments asked for after some questions, the participants highlight and give precisions on the points raised above in particular about the resources: sometimes it is one computer that is needed and not being provided. HPC resources are also cited, certainly due to the needs related to AI and DL.
Many others raise the problematic linked to lack of budget, lack of professionals to install and run data centres and difficulty to find collaborators or join collaborations to work in team.

Being asked what they would consider to improve research productivity in their country, the participants elaborated on isolation: scientists and engineers would gain a lot by working together and by having collaborations within African countries as well as with foreign countries.

## 9. Conclusion and recommendations

This survey was launched to evaluate the status of computing resources in the field of African physics research. The panel was mainly composed from participants from Africa and residing in Africa. Considering the answers, we draw the following guidelines to improve the situation and boost the scientific research in Africa:

- **Build and improve and the infrastructure**: the infrastructure should be made available and, if already existing, improved at a significant level.
    - **Network**: One essential part of the Computing situation is the access, availability and performance of the **Network**, i.e., Academic and Research Network, in Africa. Networks are vital for the access to data and information. This is not only true at the level of the universities and research centres, but even more at national and international level with connection to other countries. We need to have a global picture of the Network status in order to know the possible problems for the research groups and draw the strategy for improvement. Without filling this gap, there is no possibility of collaboration or share of knowledge. An African coordinated initiative would be a real asset.
    - **Storage and computing power** are necessary to store and process the data, which is the only way to produce results and science. The computing needed is more and more sophisticated now that Artificial Intelligence and Deep Learning have entered the game in all sciences. As suggested by some of the participants, large data centres shared within a country or with other countries within Africa would certainly be a solution that would federate the resources, decrease the costs and the disparities between universities and countries.
    - **Qualified technical staff** are necessary to deploy and run these computing resources and make them available to the physics research scientists that would not be able to deal with Cloud deployment or computer access to storage. Here a collaboration between different African countries and foreign countries could be a fruitful initiative to share IT technicians, setup few test sites, and start having an infrastructure on site.

- **Build Knowledge and include computing in Education**: The poll has highlighted the insufficient level of education in computing. Many solutions should be envisaged simultaneously:
    - Increasing the level of **computing courses** in the cursus of the physics (and other sciences) students.
    - Training **IT professionals** to prepare and operate the infrastructure.
    - Organising regular **workshops and training**. This would be highly beneficial for knowledge sharing and knowledge update to stay in the forefront in computing where evolution is very fast. But this would have an important positive side effect: Researchers have highlighted the fact that they quite often work isolated. These workshops are the best place to meet their peers and initiate collaborations that would only be beneficial to raise the research productivity.
    - Last but not least, **national and international collaboration** with others more advanced in these fields throughout the world would speed up the knowledge transfer and build collaborations that would be mutually beneficial.

> The top priority is raising the awareness of the governing bodies and stakeholders at each level: state, university, research centres, about the importance of the computing in physics research. This is absolutely necessary as this evolution needs strategic planning over years.
> Budget should be expressly dedicated to computing whether it is at the personal computer level as well as to the level of building and running large infrastructure.
> None of the main discoveries of the last decade would have been made possible without the collaborative work and the distributed use of powerful data centres all over the world.

## 10. Acknowledgments

I would like to thank all the participants that have taken some time to respond to the survey. My thanks go also to the organizers of 2021 ACP conference, to the ASFAP steering committee and to José Salt and Uli Raich for their comments and advice on this paper.
3